\documentclass[authoryear]{elsarticle}

\usepackage{lineno,hyperref}

\modulolinenumbers[5]
\usepackage{natbib}

\usepackage{amsmath,amssymb,amsfonts}
\usepackage{algorithmic}
\usepackage{graphicx}
\usepackage{textcomp}
\usepackage{xcolor}
\usepackage{enumitem}
\usepackage{float}

\begin{document}

\journal{Journal of {\it{ESWA}}}

\begin{frontmatter}



\title{Adaptive Fraud Detection System Using Dynamic Risk Features}

\author[mymainaddress,mysecondaryaddress] {Huiying Mao}
\ead{mhuiying@vt.edu}

\author[mysecondaryaddress]{Yung-wen Liu}
\ead{yungliu@microsoft.com}

\author[mysecondaryaddress]{*Yuting Jia}
\cortext[mycorrespondingauthor]{Corresponding author}
\ead{yutjia@microsoft.com}

\author[mysecondaryaddress]{Jay Nanduri}
\ead{Jay.Nanduri@microsoft.com}

\address[mymainaddress]{Department of Statistics, Virginia Tech, Blacksburg, VA 24060, USA }
\address[mysecondaryaddress]{Membership \& Knowledge Growth, Microsoft, Redmond, WA 98052, USA}

\begin{abstract}
eCommerce transaction frauds keep changing rapidly. This is the major issue that prevents eCommerce merchants having a robust machine learning model for fraudulent transactions detection. The root cause of this problem is that rapid changing fraud patterns alters underlying data generating system and causes the performance deterioration for machine learning models. This phenomenon in statistical modeling is called ``Concept Drift''.  To overcome this issue, we propose an approach which adds dynamic risk features as model inputs. Dynamic risk features are a set of features built on entity profile with fraud feedback. They are introduced to quantify the fluctuation of probability distribution of risk features from certain entity profile   caused by concept drift. In this paper, we also illustrate why this strategy can successfully handle the effect of concept drift under statistical learning framework. We also validate our approach on multiple businesses in production and have verified that the proposed dynamic model has a superior ROC curve than a static model built on the same data and training parameters.
\end{abstract}


\end{frontmatter}

\section{Introduction}\label{sec:introduction}
In the eCommerce industry, Fraud Detection Systems (FDSs) play an important role when the online fraud is increasing and spreading rapidly.  Online purchases are made on a, known as, Card Not Present (CNP) environment, where no physical cards or cardholder signatures are required. This provision of convenience, however,  also generates a fertile ground for cybercrime. In CNP scenario, fraudsters only need to provide credit/debit card information to execute a purchase. If the purchase is made successfully by a fraudster, once identified, the legitimate cardholder can file a dispute (i.e. chargeback) to the card issuing bank. For CNP transactions, merchants are also required to be responsible for the fraud financial liability. Therefore, an effective fraud detection system is essential for the eCommerce merchants. The system needs to be able to distinguish legitimate transactions from fraudulent ones and make prompt decisions (reject/approve) as required most of the time.

\begin{figure}[H]
    \centering
    \includegraphics[width=3.5in]{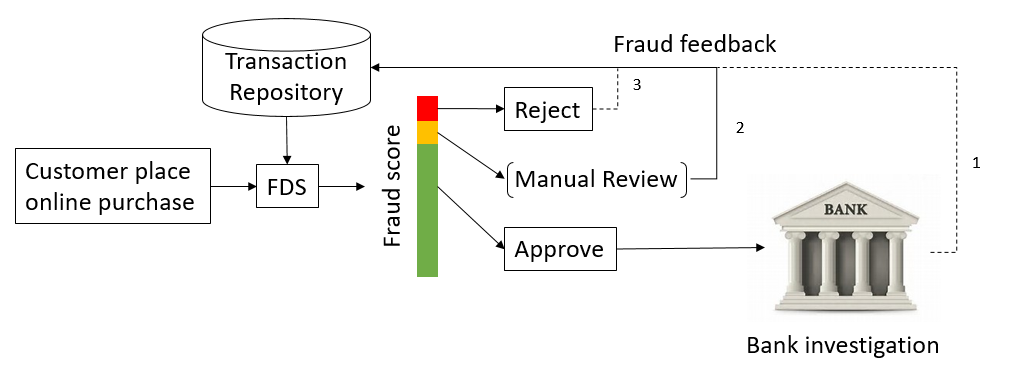}
    \caption{Mechanism flowchart of an FDS.}
    \label{fig:FDS}
\end{figure}

A simplified flowchart of a fraud detection system is depicted in Figure \ref{fig:FDS}.  Given an online purchase, a FDS can approve, reject or send the transaction to the manual review team, if applicable. The decision is made based on the risk score estimated using a machine learning (ML) model.  Different decisions result in different types of feedback, which are constantly looped into the FDS for the model performance improvement. There are three types of fraud feedback. Different types of  feedback have different delay schedules and  provide different confidence levels indicating the true status of the transaction (either legitimate or fraudulent) 

\begin{enumerate}[label=(\roman*)]
\item 
    Chargeback: this type of feedback often returns late. It could take up to several weeks or months starting from the legitimate cardholders filing the dispute to the time when merchants really receive the chargeback request. The transactions that end up with a chargeback request are often labeled as fraudulent transactions by the merchants. 
    \item 
    Manual review (MR) decisions: this type of feedback usually returns ranging from several minutes to a few hours. Although the decision made MR can be subjective, but MR rejections can be considered as a fraudulent signal with a high level of confidence. Depending on budget constraints and practical needs, not every eCommerce business has human investigators, so not every FDS can have this type of feedback.
    \item 
    System rejects: this is almost real-time from the FDS. The decision is made automatically without additional reviews. Therefore, the reliability of this type of feedback may be doubtful and need to be used more carefully. However, in practice, a customer can always escalate for the wrongful rejection by contacting the customer service. The final decisions made by customer service can be later used for the model evaluation and performance improvement. 
\end{enumerate}

The behaviors of legitimate customers and fraudsters keep evolving over time either intentionally or unintentionally.  For example, the behavior of the customers  may change abruptly due to the launch of marketing promotion, advertising, and new products.  However, the behavior of online fraudsters tend to adapt even more quickly.  As soon as online merchant adapts new strategy to prevent fraud, the online fraudsters may easily find another weakness/loophole to exploit or they could alternatively focus their attention elsewhere.  Therefore, compared with traditional classification problems, the FDSs have more challenge to overcome for eCommerce merchants. 

Due to the aforementioned scenario, the stream drawn from transactions in an eCommerce application is commonly not stationary, that is, the data are not drawn from a fixed distribution. 
This phenomenon is called ?concept drift?,  and there is  a wealth of research devoted to this topic; see
{\citep{dal2015credit, dal2017credit, Gama2014survey, gao2007general, Ditzler2015}} and references therein.  

In a non-stationary or drifting environment, a non-adaptive model that is trained under the false stationary assumption would not perform well or even fail completely at worst. There has been a need for efficient and adaptive algorithms for learning in a drifting environment since the beginning of machine learning, and now is ever increasing driven by the big-data phenomenon witnessed in the past decade.  Research work related to learning with concept drift has also been growing and many drift-aware adaptive learning algorithms have been developed. Some similar adaptive strategies have been developed independently under different names in different contexts. There are two primary families of strategies  referred to as {\emph{active}} and {\emph{passive}} approaches {\citep{Elwell2011}}. Active approaches have a detection mechanism of the changes in the data distribution, which activates an adaption once certain thresholds are reached. Passive approaches constantly update the model without requiring an explicit  detection mechanism. 

While there have been many excellent research works done on adaptive learning in a general setting, not much published works have addressed real life machine learning in the challenging eCommerce environment. We list some challenges (not inclusive) here.

\begin{enumerate}[label=(\roman*)]
 \item A decision has to be made in a blink of second (typically 20-1000 milliseconds) with high accuracy. In practice, if it takes the FDS too long to make a decision for a purchase, the system may have a timeout and generally the transaction is approved. This opens a gate for fraudsters.  If the response time is short and the decision is made with high inaccuracy, the result is costly, too. If too many fraudulent transactions are approved inline and many chargebacks are filed later, this definitely could incur a huge loss to the merchants. If too many legitimate transactions are rejected, the customers may choose to shop elsewhere or the customer service has to handle high volume of escalation calls. In the presence of manual reviews, if the review agents are flooded with a huge volume of transactions, and their review quality will suffer.  When the fraud attack happens at the worst time such during holiday season or the fraud detection system halts due to a failure in software or hardware, it would no doubt exacerbate the situation
 \item 
 The behavior of fraudster often changes very rapidly by exploiting an vulnerability of an FDS such as using stealing payment instruments from legitimate customers to  make a huge volume of purchases within a short time period. 
  
 \item For the active strategy to work, an effective and efficient detection system with rapid change in the data distribution often involves high computation cost and thus might not be justified in the revenue gained.  
\end{enumerate}

The motivation for this research stemmed from many years of handing concept drift in industry and also from the drawbacks of the current fraud detection systems. Our main contribution in this paper is to provide an effective solution to tackle the typical challenges in the eCommerce industry, that was mentioned in the previous paragraphs. Different from the existing approaches, we propose a new method that uses dynamic risk features to track the concept drift and further build a dynamic model. Our strategy is to quantify concept drift by using an entity profile with fraud feedback. This entity profile continues being updated by consuming fraud feedback signals. The statistics derived from the entity profile are then used as input features at both training and scoring stages. This approach enables the machine learning model to be self-adaptive during concept drift. Instead of focusing on model architecture and training methodology in the existing literature, we focus more on those features that can effectively adapt to concept drift in an eCommerce industry.
By keeping relatively fewer features, we can detect changes in data stream without incurring high computational cost. The results were validated in production environment and we found that models built with this approach could effectively handle concept drift. 
In essence, our proposed approach is a delicate combination of both active and passive strategies for learning in a nonstationary environment.
In addition, the experiment conducted using real data in a production setting showed that models thus built were scalable and robust.

The rest of the paper is structured as follows. Section \ref{sec:concept} provides a brief explanation of concept drift.
Section \ref{sec:survey} gives a brief summary of approaches to adaptive learning algorithms.  Section \ref{sec:model} introduces three types of dynamic risk features, and details the dynamic modeling strategy for handling concept drift. The results of applying our model to real data are shown in section \ref{sec:validation}. Section \ref{sec:conclusion} concludes this paper with a few remarks and speculation on  future research. 

\section{Concept Drift} \label{sec:concept}

In this section, we will first describe a fraud detection machine learning problem in the environment where the concept drift issue exists. Followed by specifying the root cause of concept drift, we will discuss why the existence of concept drift has some negative impact on the performance of static models. Lastly, various statistics used to track and quantify concept drift will be discussed. We refer to {\citep{Gama2014survey}} for a comprehensive survey on concept drift adaptions and to {\citep{Webb2016}} for quantitative analysis of drift.

\subsection{Definition of concept drift and two scenarios}

Concept drift is a phenomenon when the underlying data generating system changes over time {\citep{Elwell2011}}. We denote the features of a transaction as $X$ and the status of this transaction as $Y$. $X$ is a $p$-dimension random vector, where $p$ is the number of features that describe this transaction. Examples of features can be ``product name'', ``purchase dollar amount'', ``device type'', and so forth. $Y$ is a binary variable whose value is either 1 or 0, indicating the transaction status being fraudulent or legitimate.  Concept drift is said to occur when
\begin{equation}
   ||p_{t+1}(X, Y) - p_t(X, Y)|| > \delta,
\end{equation}
where $p_t$ represents the joint probability density function (PDF) of $X$ and $Y$ at time $t$, time $t+1$ refers to the next timestamp,  $||\cdot||$ is a norm to measure the difference between the two density functions, and $\delta$ is a pre-defined threshold. Such a data generating system with underlying concept drift is also referred as non-stationary environment (NSE) in the literature (e.g., {\citep{ditzler2013incremental}}).

We use $p(X|Y)$, a conditional PDF that gives the conditional probability of $X$ given $Y$,  to describe the likelihood of presence of transaction features. For purchases made by good customers, the distribution of transaction features is $p(X|Y=0)$. Similarly, the distribution of transaction features from fraudsters is described by $p(X | Y=1)$. Because of
\begin{equation}
    p(X, Y) = p(X|Y)p(Y),
\end{equation}
the fluctuation of joint PDF in (2) is originated from the fluctuations of $p(Y)$ and $p(X|Y)$. We can divide concept drift into two scenarios:
\begin{enumerate}[label=(\roman*)]
    \item 
    $p_{t+1} (Y) \neq p_t (Y)$: the overall population distribution shifts between good customers and fraudsters at the different time points?$t$ and $t+1$. For example, under economic prosperity at time $t$, higher portion of purchases are made by good users compared with the regular economic status at time $t+1$ . For another example, fraudsters test the FDS constantly. When they find a loophole in the system, they would place much more orders. As a result, the percentage of fraudulent transactions jumps. In other words, the good/bad purchase volume distribution change leads to the variation in $p_t (Y)$ with respect to time $t$.
    \item 
    $p_{t+1} (X|Y=y) \neq p_t (X|Y=y )$: the shopping features of good customers $p(X|Y=0)$  or fraudsters $p(X|Y=1)$ at time $t+1$ differs from their shopping features at time $t$. For instance, customers adjust their purchase behavior from product $A$ to product $B$ after an advertising campaign; fraudsters heavily attack a certain product when the product is popular in the market.
\end{enumerate}
These two scenarios are not necessarily mutually exclusive. They often occur simultaneously.

\subsection{Performance degradation of static model}\label{subsec:2drift}

Supervised learning models, where the status of the subject is known, are commonly applied by a FDS.  Classified and Regression Trees (CART), Neural Network model, and Support Vector Machine (SVM) are the most commonly used classifiers . In supervised learning, the properties of conditional PDF $p(Y|X)$ is of the most interest {\citep{friedman2001elements}}. Under concept drift, $p(Y|X)$ may or may not change based on
\begin{equation}
    p(Y|X) = \frac{p(X,Y)}{p(X)},
\end{equation}
Literature categorizes concept drift into ``real'' and ``virtual'' versions based on the existence of fluctuation in $p(Y|X)$; see {\citep{Gama2014survey,gao2007general,hoens2012learning,kelly1999impact}}:
\begin{enumerate}[label=(\roman*)]
    \item Real concept drift if $p_{t+1} ( Y | X ) \neq p_t ( Y | X)$.
    \item Virtual drift if $p_{t+1} (Y | X ) = p_t ( Y | X )$.
\end{enumerate}
This paper focuses on handling real concept drift. In the following context, unless specified, ?concept drift? refers to ``real concept drift''.

The objective of having machine learning model in a fraud detection system is that to use this model $\mathcal{M}$ to effectively estimate the conditional probability, i.e.
\begin{equation}
    \mathcal{M}(X) \propto p(Y=1|X).
\end{equation}
$\mathcal{M}$ outputs a score or a probability, which measures the risk level of the transaction. That is, given enough volume of stationary historical transactions, a sophisticated learning algorithm is used to estimate the probability of a transaction with features $x$ being fraudulent $P( Y = 1|X = x)$, where $x$ is generated from the same stationary environment. For notation simplicity, from now on $P(Y =1 | X = x )$ will be denoted as $p_x(Y=1)$.

When concept drift happens, without taking this factor into account, the probability of a transaction with features $x$ generated at time $t+1$ being fraudulent, $p_{x, t+1} (Y=1)$, is wrongly predicted by $p_{x, t} (Y=1)$. In other words, suppose $\mathcal{M}_t$ is a model trained based on historical data with fraud labels up to time $t$:
\begin{equation}
    \{ \mathcal{Y}_t, \mathcal{X}_t\},
\end{equation}
where $\mathcal{X}_t$ represents the set of all transactions and $\mathcal{Y}_t$ represents the set of corresponding labels. For a new transaction with feature $x$ generated at time $t+1$,
\begin{equation}
\mathcal{M}_t(x)\propto p_{x,t}(Y=1) \neq p_{x, t+1}(Y=1).
\end{equation}

The inequality is the root cause of incorrect prediction. In this paper, such a model without concept drift handling strategy is referred as ``static model''.

\begin{figure}[H]
    \centering
    \includegraphics[width=3in]{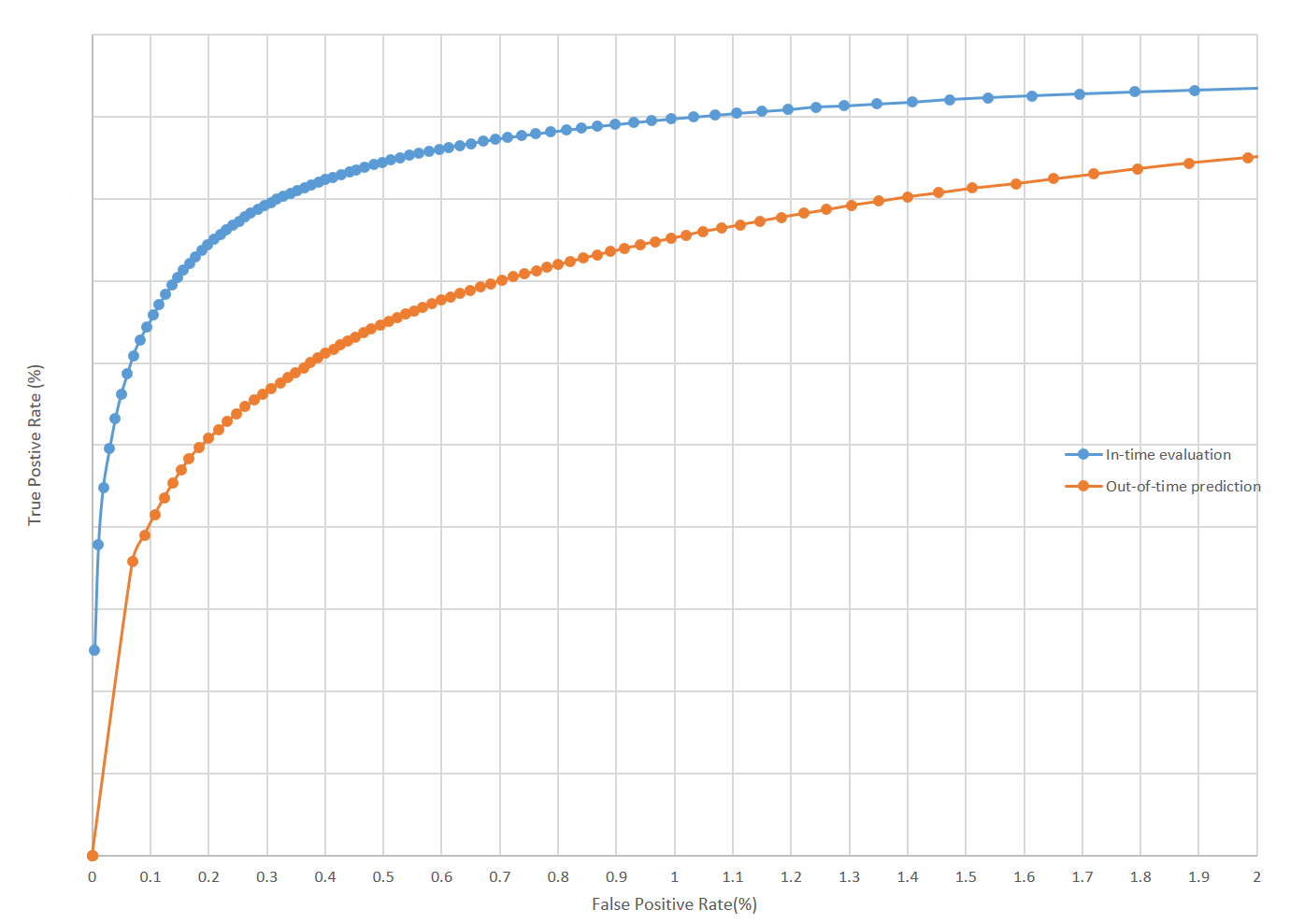}
    \caption{Performance degradation of static model.}
    \label{fig:perf_deg}
\end{figure}

Due to concept drift, model performs much worse than expected. ROC curve in Figure \ref{fig:perf_deg} shows an example of comparison between in-time prediction (blue) and offline prediction (orange) from one of our business portfolios. In-time means that model performance is evaluated on a dataset that shares the same time range with training dataset; offline refers to that the test dataset is collected outside the training dataset time range.

\section{Survey of Adaptive Algorithms}\label{sec:survey}

In this section, we give a brief summary of approaches and algorithms that address concept drift issues which are mostly relevant to this paper. We recommend articles {\citep{Tsymbal2004, Gama2014survey, Ditzler2015}} for a comprehensive treatment.  

Fraud detection systems in the eCommerce use adaptive learning algorithms to process high-speed flows of stream data. 
Based on whether there is a mechanism to detect change in data distribution, adaptive algorithms for learning in the presence of concept drift are primary based on \emph{active} or \emph{passive} approach. Active algorithms detect concept drift, and passive algorithms  constantly update the model with new data, regardless whether drift is present. 

In an active approach that handles concept drift, there are two phrases: change detection and adaption. The change detection mechanism rarely operates 
directly on raw data, but instead on independent features that are often based domain expert knowledge, and are extracted from incoming data stream.  Once a change is detected, the classifier needs to adapt to the change from the newly available information and to discard obsolete ones. The adaption can be done through windowing, weighting  or random sampling {\citep{Ditzler2015}}.   

In the literature, two main approaches for change detection had been proposed and adopted, which differ for the different entity in analysis: based on form of distributions of the input data, and based on the classification error. The first approach detects the change in the jointed pdf structure. The second approach evaluates variations in the classification error on supervised data, and the classification accuracy uses a fixed threshold or an adaptive one {\citep{Gamma2004}}. In both cases, a change is detected as soon as the classifier's accuracy falls below the threshold. The first approach assesses the drift of the pdf of the inputs disregarding their label values {\citep{Alippi2008}}.  

A combination of both approaches was proposed in {\citep{Alippi2012}}. The proposed solution assesses  the stationary in both the joint
probability density function of the labeled data and the distribution of the inputs on unlabeled data. 
Dal Pozzolo \textit{et al.} treated immediate feedback samples and delayed samples, whose labels are obtained only after some time, separately. They suggested to trained two distinct classifiers based on each type of feedback respectively, and then aggregate the outputs {\citep{dal2015credit}}.  
Gao \textit{et al.} proposed a method which uses an ensemble of classifiers built on sequential chucks of training samples to handle concept drift. In their approach, each classifier is trained on a short period of time and updated frequently {\citep{gao2007general}}.  

\section{Model with dynamic risk features}\label{sec:model}

In this paper, we propose a dynamic model that can overcome the model performance degradation caused by concept drift. Our strategy is to incorporate concept drift measures as dynamic risk features into model training and scoring. By design, the model learns not only from the original static features but also from dynamic risk features, which provide an effective measurement of concept drift.  As a result, the model \emph{self-adapts} when concept drifts happen. This approach saves the effort to constantly retrain the model for preventing the performance degradation.

\subsection{Measuring concept drift using entity profile with fraud feedback}

As shown in Section \ref{sec:concept}, concept drift can be measured using the variation on $p(Y)$, $p(X|Y)$ and $p(Y|X)$. Therefore, to track concept drift, we need to consistently monitor the probability distributions with respect to time $t$ and to estimated $p_t (Y)$, $p_t (X|Y)$, $p_t (Y|X)$. However, the estimates of $p_t (Y)$, $p_t (X|Y)$, $p_t (Y|X)$ are not the adequate measurement for concept drift since they are the instantaneous measures and could vary from the norm most of them time. Therefore, in our study, we use the average probabilities within a sliding window as surrogates, and  the average probabilities are approximated by frequency ratios.

To formulate the methodology: let $p_t=p(t)$ denotes the probability of interest at time $t$, which changes over time. The average probability $\bar p_{t,h}$, the surrogate of $p(t)$, is defined as
\begin{equation}
    \bar p_{t,h} = \frac{\int_{t-h}^t p(t) dt}{h},
\end{equation}
where $h$ is the length of a sliding window. The $\bar p_{t,h}$ is approximated through some statistics $f_{t,h}$ by the law of large numbers. That is,
\begin{equation} \label{eq:f_pbar_p}
    f_{t,h} \approx \bar p_{t,h}, \  \lim_{h \rightarrow 0} \bar p_{t,h} = p_t,
\end{equation}
and thus
\begin{equation} \label{eq:ft_to_pt}
    \lim_{h \rightarrow 0} f_{t,h} \approx p_t.
\end{equation}

The length of the sliding window needs to be chosen carefully. It should be long enough to have a good sample size for reliable estimation, while short enough to be sensitive to the fluctuation. Our approach is to use two sliding windows with  lengths of $h$ and $H$ ($h<H$). That is, at time $t$, $\hat{f}$ statistics are calculated based on the transactions happened within the short$t$-term time span $[t-h,t)$ and long-term time span $[t-H,t)$, respectively. Later discussion will illustrate the statistics calculation only on short-term time span, as that for the long-term span is very similar.  

\subsubsection{Using overall fraud rate (FR) to measure $p_t (Y)$}

By definition,
\begin{equation}
    p_t(Y) = 
            \begin{cases} 
              P(Y=1|t) & \mbox{if } Y=1, \\
              1-P(Y=1|t) & \mbox{if } Y=0; \\
            \end{cases}
\end{equation}
and $P(Y=1|t)$ is estimated by overall fraud rate (FR) within the time window $[t-h,t)$, which is defined as
\begin{equation}
\begin{split}
      &\mbox{overall FR}_t = \\ 
      &\frac{\mbox{\# of fraudulent transactions within time window } [t-h,t)}{\mbox{\# of total transactions within time window } [t-h,t)}.
\end{split}
\end{equation}
Besides overall $\mbox{FR}_t$, overall dollar-weighted fraud rate (\$FR) is also a suitable candidate to be included as a dynamic risk feature:
\begin{equation}
\begin{split}
      &\mbox{overall \$FR}_t = \\ 
      &\frac{\mbox{\$ of fraudulent transactions within time window } [t-h,t)}{\mbox{\$ of total transactions within time window } [t-h,t)} .
\end{split}
\end{equation}
Overall $\mbox{\$FR}_t$ does not directly estimate $P(Y=1|t)$, but it puts more weights on higher dollar value transactions and helps model to learn the fraud pattern change.
\begin{figure}[H]
    \centering
    \includegraphics[width=3.5in]{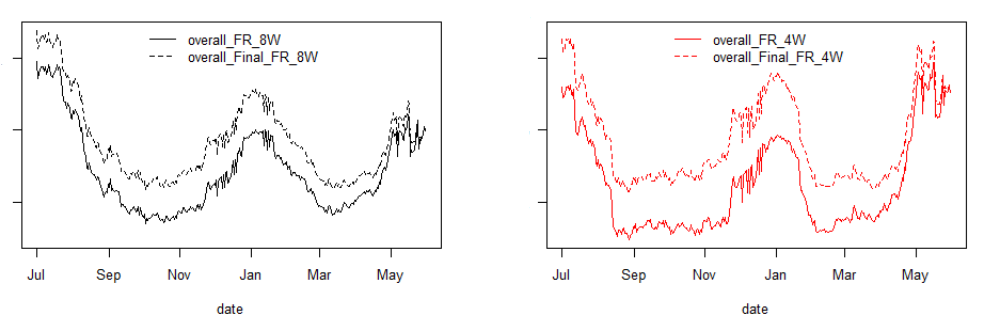}
    \caption{Overall fraud rate with respect to time for one business partner.}
    \label{fig:FR}
\end{figure}

Figure \ref{fig:FR} shows overall fraud rate calculated on the long-term sliding window (left) and short-term sliding window (right).  The solid lines are calculated at time $t$. There is a varied time delay before receiving the chargeback signal. We also draw the final fraud rates (dash lines), which are calculated offline when this study was conducted,  for comparison. These two lines follow the same trending pattern. This justifies Formula (\ref{eq:f_pbar_p}). The difference between the solid and dashed lines diminishes because that more recent transactions have higher percentage of chargebacks having not been received yet when plotting them.

\subsubsection{Using fraud rate of each transaction entity $\mbox{FR}_t(E)$ to measure $p_t (Y|X)$}
In practice, a transaction entity $E_X$ is often described  by a set of features, $X$. Therefore, we propose measuring the fraud rates profiled on important features. Suppose $E_X$ is the transaction entity with associated features:  $X = \{x_1,x_2,\cdots \}$. Within the time window  $[t-h,t)$, the number of good and bad transactions and their dollar amounts for each transaction entity and marginal totals are shown in Table \ref{tab:entity_profile}.

\begin{table}[H]
\caption{Transaction count and dollar amount distribution profiled on entity $E_X$ within time window  $[t-h,t)$}
\label{tab:entity_profile}
\begin{center}
\renewcommand{\arraystretch}{2}
\begin{tabular}{|c|ccc|c|}
\hline
 & $x_1$ & $x_2$ & $\cdots$ & Total \\
\hline
Bad(Y=1) & $N_{11} (\$_{11})$ & $N_{21} (\$_{21})$ & $\cdots$ & $N_{\cdot1} (\$_{\cdot1})$ \\ 
Good(Y=0) & $N_{10} (\$_{10})$ & $N_{20} (\$_{20})$ & $\cdots$ & $N_{\cdot0} (\$_{\cdot0})$ \\ 
\hline
Total & $N_{1\cdot} (\$_{1\cdot})$ & $N_{2\cdot} (\$_{2\cdot})$ & $\cdots$ & \\
\hline
\end{tabular}
\end{center}
\end{table}

$N_{i1} (\$_{i1})$ represents the number (dollar amount) of fraudulent transaction with feature  $x_i, i=1,2,\cdots$;  $N_{i0} (\$_{10})$, within time window $[t-h,t)$. $N_{i0} (\$_{i0})$ are for the number (dollar amount) of good transactions with feature $x_i, i=1,2,\cdots$, within the sample period.  Their marginal totals are calculated by
\begin{equation}
    N_{\cdot j} = \sum_i N_{ij}, j = 0,1; N_{i \cdot} = N_{i0} + N_{i1}, i = 1,2, \cdots
\end{equation}
\begin{equation}
    \$_{\cdot j} = \sum_i \$_{ij}, j = 0,1; \$_{i \cdot} = \$_{i0} + \$_{i1}, i = 1,2, \cdots
\end{equation}
Thus, the fraud rates and dollar weighted fraud rates for entity $E$ within time window $[t-h,t)$ are:
\begin{equation}
    \mbox{FR}_t(x_i) = \frac{N_{i1}}{N_{i\cdot}},\quad i = 1,2, \cdots
\end{equation}
\begin{equation}
    \mbox{\$FR}_t(x_i) = \frac{\$_{i1}}{\$_{i\cdot}},\quad i = 1,2, \cdots
\end{equation}
They are the risk features used to measure $p_t (Y|X)$.

Figure 4 shows the fraud rates of a selected object. For example, we can select ?product name? as the entity and one particular product as the selected object. Red line is the fraud rate calculated by 4-week sliding windows and the black line is for 8-week sliding windows. As we can see, this object is under attack around the November 2016. FR calculated within the 4-week sliding window is more sensitive to fraud attack. When the short-term FR is higher than long-term FR, it indicates that fraudsters are attacking this product recently. On the other hand, if short-term FR is lower, it means fraudsters divert their attention to the other products. Also, the difference between long-term FR and short-term FR reflects the fraud attack severity.

\begin{figure}[H]
    \centering
    \includegraphics[width=3.5in]{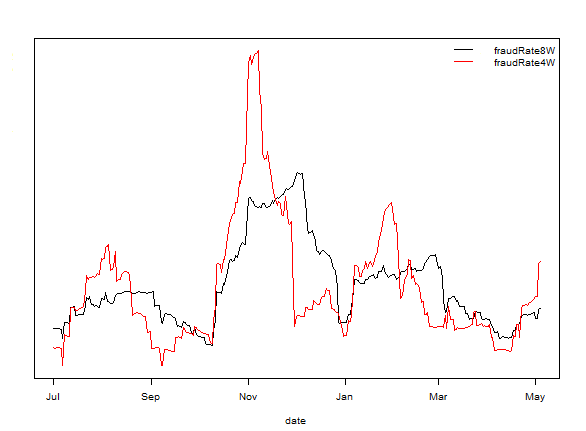}
    \caption{Entity profiled fraud rate for a selected object.}
    \label{fig:individual_FR}
\end{figure}

\subsubsection{Using Weight of Evidence on features of entities, (WoE($X_{E_\ell}$)),  $\ell = 1,2,\dots$ to measure $p_t (X|Y)$}

Similarly,
\begin{equation}
    p_t (X|Y) = 
    \begin{cases} 
      P(X|Y=1, t) & \mbox{if } Y = 1, \\
      P(X|Y=0, t) & \mbox{if } Y = 0; \\
    \end{cases}
\end{equation}
and $X$ is high-dimensional. We propose to use the estimation of 
\begin{equation}
    \ln\Big[ \frac{p_t(X_{E_\ell}|Y=1)}{p_t(X_{E_\ell}|Y=0)}\Big]
\end{equation}
to measure $p_t (X | Y)$. The estimation is known as \textit{Weight of Evidence (WoE)} in credit risk modeling {\citep{anderson2007credit}}, formulated by
\begin{equation}
    \mbox{WoE}_t(x_i) = \ln \Big[ \frac{N_{i1}/N_{\cdot 1}}{N_{i0}/N_{\cdot 0}} \Big], \quad i=1,2,\cdots,
\end{equation}
\begin{equation}
    \mbox{\$ WoE}_t(x_i) = \ln \Big[ \frac{\$_{i1}/\$_{\cdot 1}}{\$_{i0}/\$_{\cdot 0}} \Big], \quad i=1,2,\cdots,
\end{equation}
where $N_{i1} (\$_{i1} ),N_{i0} (\$_{10})$ are the count and dollar amount for bad and good transactions happened within time window $[t-h,t)$ as shown in Table \ref{tab:entity_profile}. 

For a particular entity feature value $x_i, \mbox{WoE}(x_i)$ is the difference between the log of odds ratios of $x_i$ and that of the overall population:
\begin{equation}
    WoE(x_i) = \ln \Big[ \frac{N_{i1}/N_{\cdot 1}}{N_{i0}/N_{\cdot 0}} \Big] 
             = \ln \Big[ \frac{N_{i1}}{N_{i0}} \Big] - \ln \Big[ \frac{N_{\cdot 1}}{N_{\cdot 0}} \Big].
\end{equation}

Therefore, when WoE$(x_i) > 0$, the odds of $x_i$ being fraudulent feature is higher than the average; when WoE$(x_i) < 0$, the odds of $x_i$ being fraudulent feature is lower compared to the average. That is to say, WoE provides an indication of fraudsters? attacking target feature. 

\begin{figure}[H]
    \centering
    \includegraphics[width=3.5in]{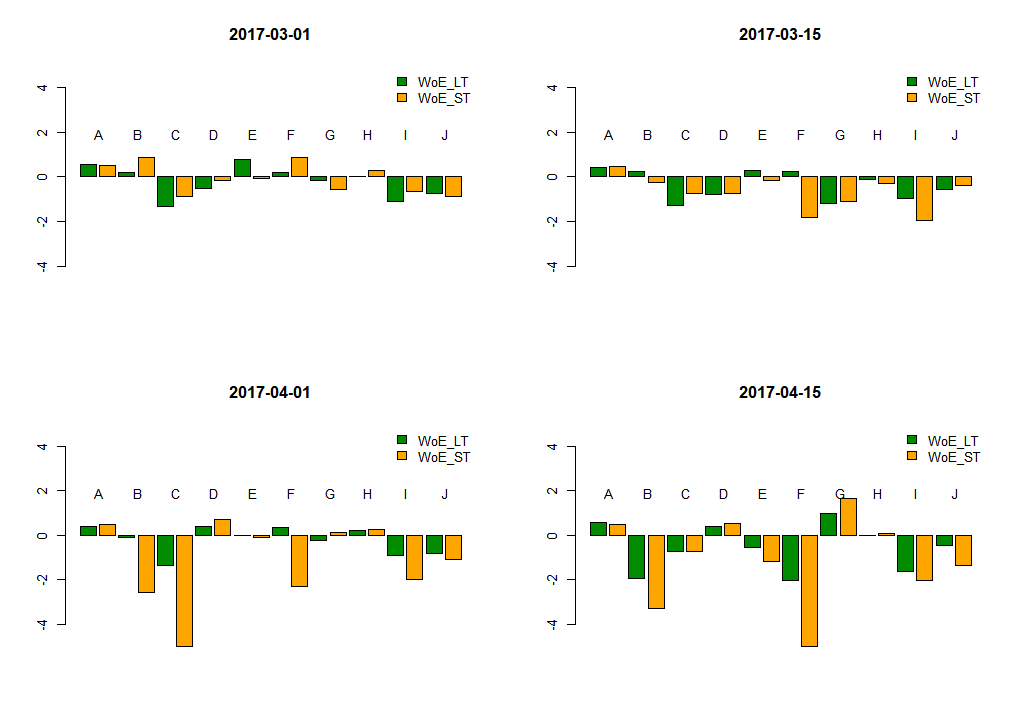}
    \caption{Snapshots of long-term and short-term WoE distribution for 10 products.}
    \label{fig:WoE}
\end{figure}

For example, in Figure \ref{fig:WoE}, among 10 products A-J in the market:  Product B is targeted by fraudsters on March 1st where its WoE is above zero. But as time goes by, Product B is no longer the attack object; instead, Product G becomes the attack target on April 15.

One final note for this subsection is that the selection of features is pivotal. Important features could include  product name, account email domain, billing country, merchant name, and so forth. An feature should be selected such that the number of feature values should not be too few or too many. If too few, the feature value is too broad to have the probability fluctuation to show up. For example, if ``market country'' is chosen as the feature, then fraudster attack different product within the same country won?t appear in the tracking. Likewise, it should not have too many feature values. Otherwise, the frequency ratio approximations for each feature would be  unstable.

\subsection{Measuring concept drift using entity profile with fraud feedback}

So far, we have shown multiple tracking statistics which can provide comprehensive description to concept drift. In this subsection, we will show how to use these statistics to construct a set of dynamic risk features and use those features as model inputs.

Suppose $\{E_\ell(X) , \ell(X)=1,2,\cdots \}$ is the set of entities chosen. The dynamic risk feature set is constructed as
\begin{equation}
    F_t = 
        \left\{
          \begin{array}{@{}c@{}}
          \mbox{overall FR}_{t,h}, \mbox{FR}_{t,h}, \mbox{FR}_{t,H}, \mbox{FR}_{t,H} \\
          \mbox{FR}_{t,h}(E_\ell(X)),  \mbox{FR}_{t,h}(E_\ell(X))  \\
          \mbox{FR}_{t,H}(E_\ell(X)),  \mbox{FR}_{t,H}(E_\ell(X))  \\
          \mbox{WoE}_{t,h}(E_\ell(X)),  \mbox{WoE}_{t,h}(E_\ell(X)) \\
          \mbox{WoE}_{t,H}(E_\ell(X)),  \mbox{WoE}_{t,H}(E_\ell(X))  \\
          \ell = 1,2,\cdots
          \end{array}
        \right\}.
\end{equation}

\begin{figure}[H]
    \centering
    \includegraphics[width=3.5in]{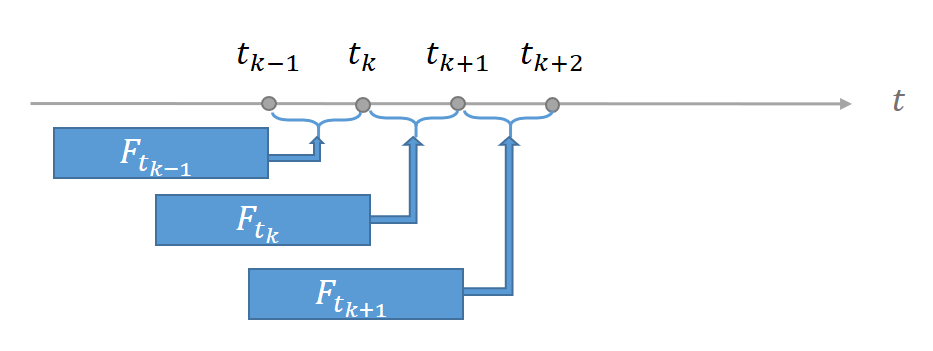}
    \caption{Attaching logic of assembling transactions.}
    \label{fig:assemble_logic}
\end{figure}

To approximate real-time fluctuation, dynamic risk features need to be updated relative frequently. Denote the pre-defined updating time stamps for dynamic risk features as $\{t_k : k \in N\}$. At every time stamp $t_k$, calculate dynamic risk feature set $F_{t_k}$ based on the transactions happened within the sliding windows $(t_{k-h},t_k)$ and $(t_{k-H},t_{k})$. Later, for transactions happened within time span $[t_k,t_{k+1})$, $F_{t_k}$ become part of the transactions attributes based on the entity, as illustrated in Figure \ref{fig:assemble_logic}.

Therefore, for a transaction with a feature $x$ happens at time $t$, the FDS provides a set of associated dynamic risk features. Since these features are calculated in a relative frequent fashion (difference between $t_{k+1}$ and $t_k$ is small), we can use $f_t$ and $f_{t_k}$ indistinguishably to denote the dynamic risk features associated with transaction feature $x$. Denote the assembled transaction as
\begin{equation}
    x' = (x, f_t)
\end{equation}
where $x'$ is $(p+q)$-dimensional. So $p$ is the number of original features and $q$ is the number of dynamic risk feature. $x$ and $x'$ refer to the same transaction. The only difference is $x'$ has both dynamic risk feature and static feature describing it, while $x$ only has static features.

Let $F_t$ be the collection of dynamic risk features calculated prior to time $t$, i.e.
\begin{equation}
    \mathcal{F}_t = \{ F_{t_k}: t_k < t\}.
\end{equation}

Denote $\widetilde{\mathcal{M}}_t$ as the model applied in production based on the training dataset 
\begin{equation}
    \{ \mathcal{Y}_t, \mathcal{X}_t, \mathcal{F}_t\}.
\end{equation}

We call $\widetilde{\mathcal{M}}_t$ a ``dynamic model'' since it includes the dynamic risk features $F_t$ as training inputs.
At time $t+1$, for transaction $x$, the output score from the dynamic model, because of assembling $f_{t+1}$  as inputs, satisfies
\begin{equation} \label{eq:Mt}
	\widetilde{\mathcal{M}}_t (x' )=\widetilde{\mathcal{M}}_t (x,f_{t+1} ) \propto p_{t+1} (x).
\end{equation}
The justification for Equation (\ref{eq:Mt}) comes from the indicating power of $f$ to $p$ as shown in Equation (\ref{eq:ft_to_pt}).

By this approach, without having to update the existing model between time $t$ and $t+1$, the dynamic model can overcome concept drift phenomenon to predict the probability of a transaction being fraudulent at time $t+1$ more accurately.

\section{Model Validation Using Real Data} \label{sec:validation}

Online purchase transaction data from two of Microsoft's business partners were used to valid our approach. Within each, we randomly select 70\% transactions for training and the rest 30\% for testing. For the same business partner data, static and dynamic models are trained over the same set of transactions and evaluated on the same test dataset. FastTree, known as an efficient implementation of the MART gradient boosting algorithm [2], is used as our model training algorithm. 
The configuration for dynamic model is
\begin{enumerate}[label=(\roman*)]
    \item Entity: product name, account email domain, billing country code, device type + currency + the first three digits/letters of a SKU;
    \item Sliding window size: 4 weeks and 8 weeks;
    \item Dynamic feature update frequency: daily. 
\end{enumerate}

\begin{figure}[H]
    \centering
    \includegraphics[width=3.5in]{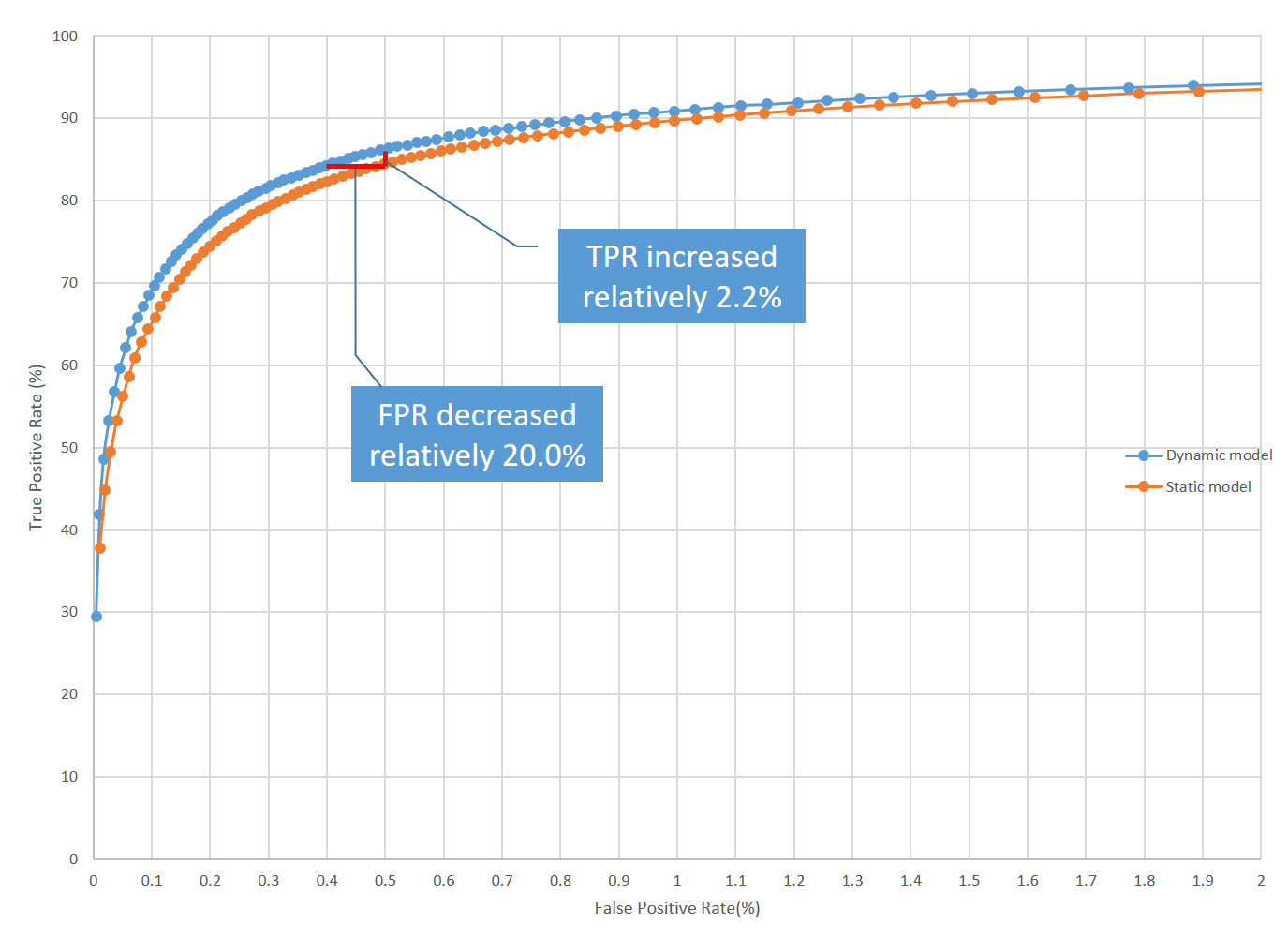}
    \caption{ROC curve comparison for partner 1.}
    \label{fig:MS_ROC}
\end{figure}

\begin{figure}[H]
    \centering
    \includegraphics[width=3.5in]{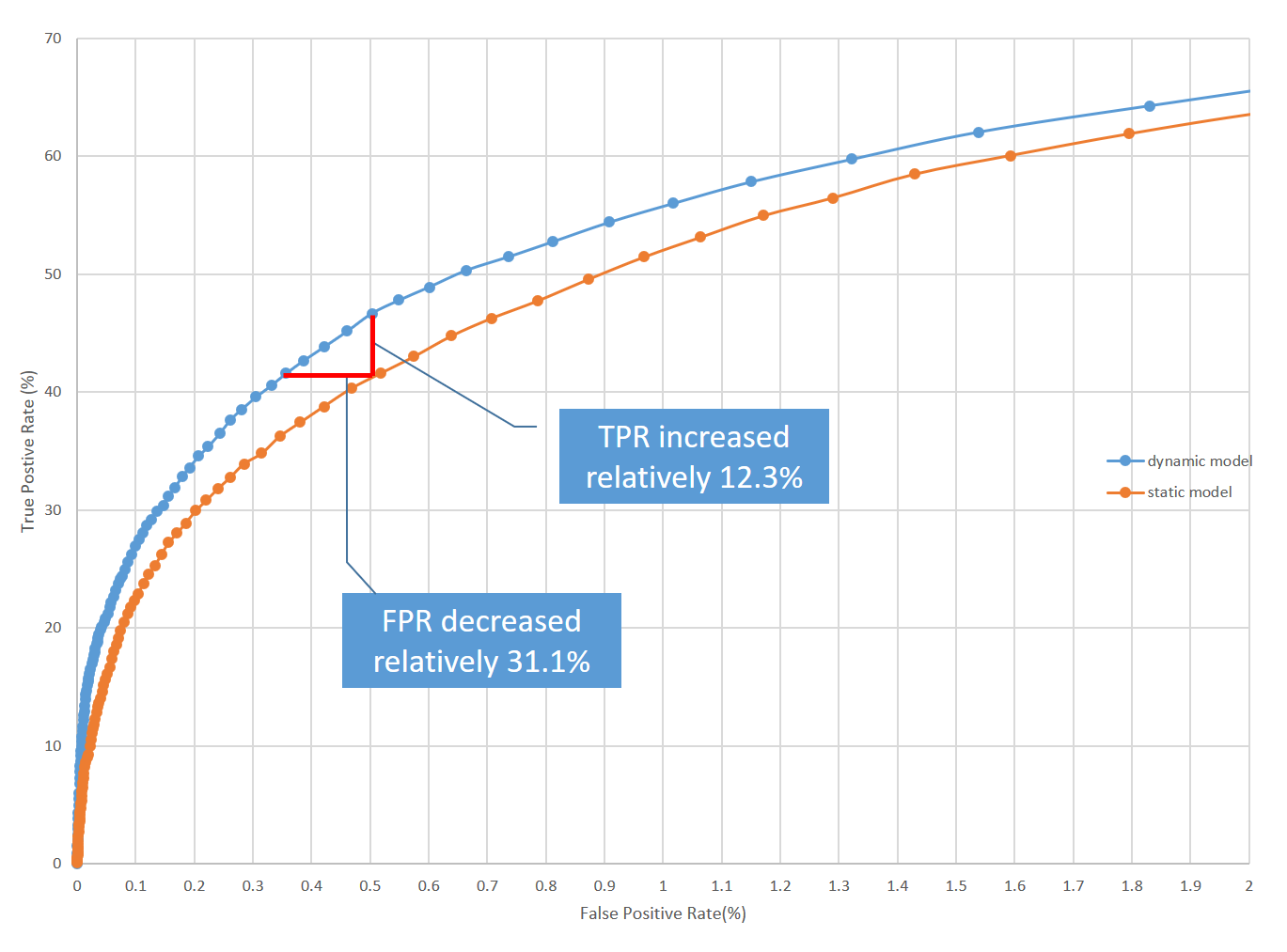}
    \caption{ROC curve comparison for partner 2.}
    \label{fig:Xbox_ROC}
\end{figure}

Figure \ref{fig:MS_ROC} and Figure \ref{fig:Xbox_ROC} show the ROC curve comparison between static model and dynamic model on business partner 1 and business partner 2 respectively. ROC curves were generated from test dataset. Blue line represents the results of the dynamic model, while orange line represents static model. These lifts confirm that dynamic model is superior for both business partners. For partner 1, while keeping false positive rate (FPR) at 0.5\% for both models, dynamic model can increase true positive rate (TPR) by 2.23\% relatively; if TPR is controlled at 84.4\%, dynamic model can reduce FPR by about 20\% relatively. For partner 2, while keeping FPR at 0.5\%, dynamic model can increase TPR by about 12.3\% relatively; if TPR is controlled at 41.6\%, dynamic model can reduce FPR by about 31.1\% relatively. We also found that dynamic risk features were shown as top features with higher information values.

\begin{figure}[H]
    \centering
    \includegraphics[width=3.5in]{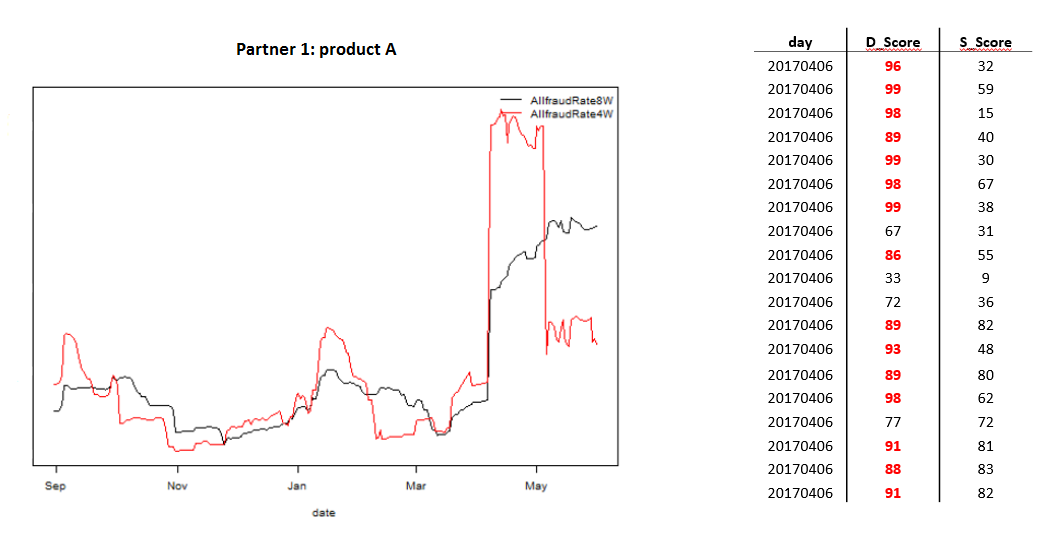}
    \caption{Fraudulent transactions dynamic model catches.}
    \label{fig:example}
\end{figure}

Figure \ref{fig:example} shows the fraudulent transactions that dynamic model caught while the static model missed. The red curve is fraud rate on one entity value calculated on the 4-week sliding windows, and black curve is for the 8-week sliding windows. The system was under fraud attack on 04/06/2017 when the red line is significantly higher. Some fraudulent transactions are shown in the right table with their corresponding scores. D\_Score is the score given by the dynamic model and S\_Score is given by the static model. If we have used the dynamic score and the cutoff as 85, we could have caught most frauds which static model missed. 

\begin{figure}[H]
    \centering
    \includegraphics[width=3.5in]{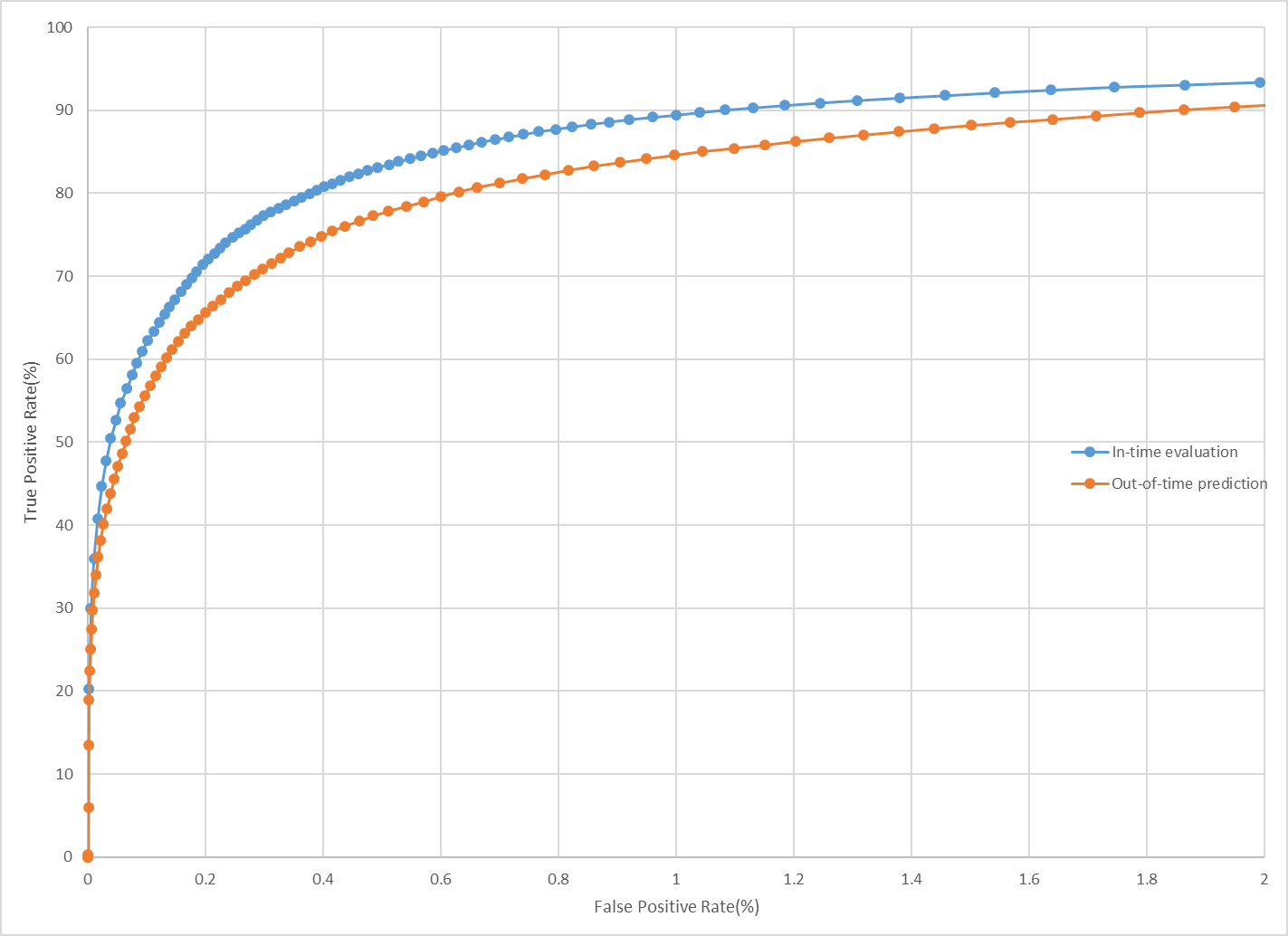}
    \caption{Performance degradation of dynamic model.}
    \label{fig:perf_deg_dynamic}
\end{figure}

Back to Figure \ref{fig:perf_deg}, due to concept drift, static model performed much worse than expected on offline data. On similar dataset, Figure \ref{fig:perf_deg_dynamic} shows the degradation of model performance is much smaller for the comparison between in-time prediction and offline validations. 
\section{Discussion}\label{sec:conclusion}

Entity profile with fraud feedback is a critical feature which brings most recent fraud patterns for dynamic models to consume. Selection of 
features for the entity profile is very crucial.
 This selection  can be treated as a problem of partitioning the data space. If, for example, product title is the chosen entity, the data space will be cut into pieces by the number of products. Since FR($X$) and WoE($X$) are calculated for each piece, the partition of data space should not be too coarse or too fine. If cutting too coarsely, concept drift won't be captured. For example, if we choose country as the major feature for entity, it will cut the data space into USA part, Mexico part, China part, etc. The dynamic risk features calculated on the countries will not reflect fraud target shifting within the countries. If cutting too finely, the estimates of FR and WoE will not be stable. How to wisely divide sample space is an interesting topic that is worth further research. 

The selection of moving window size is another important factor to consider. We selected the long-term short-term sliding windows based on experience. It works well but finding a systematic algorithm with an automatic  window selection can be an interesting topic.
 We have explored long term and short term cascade modeling which  helps to build a solid model to catch new emerging fraud patterns without impacting the detection power on existing ones.  

We mainly use the ROC curves for validation of our dynamic approach to adaptive learning, as they are often used for evaluating binary classifiers.
Other measures that related to revenue gained, or prevented losses could be interesting from a business point of view. However, these measures may not be comparable for different businesses.   


As we mentioned in the introduction section, our approach in this paper is a combination of active and passive strategies in handing concept drift for adaptive learning. We select dynamical risk features to build a light-weight detection mechanism for fraud pattern changes using signals from the input and out signals. The approach is validated in real operational settings and is proved to be effective; cf.[Section 5] {\citep{Gama2014survey}}.  Although this paper focuses on handling real concept drift as said in Subsection \ref{subsec:2drift}, our approach works for virtual drift as well.

\section*{References}


\bibliography{drfbib}
\bibliographystyle{apa}


\end{document}